\pdfoutput=1
\documentclass[%
superscriptaddress,10pt, 
preprint,
 amsmath,amssymb,
aps,
]{revtex4-1}

\usepackage{graphicx}
\usepackage{dcolumn}
\usepackage{bm}
\usepackage{xspace}
\usepackage[cdot,mediumqspace,amssymb]{SIunits} 
\usepackage[disable]{todonotes} 

\usepackage[colorlinks=true,
linkcolor=black,
urlcolor=blue,
citecolor=black]{hyperref}

\usepackage[
margin=0.75in,
]{geometry}

\newcommand{\BST}[0]{(Bi,Sb)${}_{2}$Te${}_{3}$\xspace}
\newcommand{\STO}[0]{SrTiO${}_{3}$\xspace}

\newcommand{\TK}[0]{${\theta}_{K}$\xspace}
\newcommand{\Rxx}[0]{$R_{xx}$\xspace}
\newcommand{\Rxy}[0]{$R_{xy}$\xspace}

\newcommand{\um}[0]{\micro\meter\xspace}
\newcommand{\uW}[0]{\micro\watt\xspace}

\begin{abstract}
	Many proposed experiments involving topological insulators (TIs) require spatial control over time-reversal symmetry and chemical potential. We demonstrate reconfigurable micron-scale optical control of both magnetization (which breaks time-reversal symmetry) and chemical potential in ferromagnetic thin films of Cr-\BST grown on \STO. By optically modulating the coercivity of the films, we write and erase arbitrary patterns in their remanent magnetization, which we then image with Kerr microscopy. Additionally, by optically manipulating a space charge layer in the underlying \STO substrates, we control the local chemical potential of the films. This optical gating effect allows us to write and erase \textit{p-n} junctions in the films, which we study with photocurrent microscopy. Both effects are persistent and may be patterned and imaged independently on a few-micron scale. Dynamic optical control over both magnetization and chemical potential of a TI may be useful in efforts to understand and control the edge states predicted at magnetic domain walls in quantum anomalous Hall insulators.

\end{abstract}

\begin{document}


\title{Local optical control of ferromagnetism and chemical potential in a topological insulator}

\author{Andrew L. Yeats}
\affiliation{Institute for Molecular Engineering, University of Chicago, Chicago, IL 60637.}
\affiliation{Materials Science Division, Argonne National Laboratory, Argonne, IL 60439.}
\author{Peter J. Mintun}
\affiliation{Institute for Molecular Engineering, University of Chicago, Chicago, IL 60637.}
\author{Yu Pan}
\affiliation{Materials Research Institute and Department of Physics, The Pennsylvania State University, University Park PA 16802.}
\author{Anthony Richardella}
\affiliation{Materials Research Institute and Department of Physics, The Pennsylvania State University, University Park PA 16802.}
\author{Bob B. Buckley}
\affiliation{Institute for Molecular Engineering, University of Chicago, Chicago, IL 60637.}
\author{Nitin Samarth}
\affiliation{Materials Research Institute and Department of Physics, The Pennsylvania State University, University Park PA 16802.}
\author{David D. Awschalom}
\email{awsch@uchicago.edu}
\affiliation{Institute for Molecular Engineering, University of Chicago, Chicago, IL 60637.}
\affiliation{Materials Science Division, Argonne National Laboratory, Argonne, IL 60439.}

\date{\today}

\maketitle


\section{Introduction}

The unusual topology of electronic bands in some materials can produce quantum states that are stabilized by symmetry. Topological insulators (TIs) have attracted particular attention for their symmetry-protected surface and edge states, which may hold promise for applications in spintronics and quantum computing \cite{Hasan2010, Qi2011a}. Ferromagnetic TIs combine a topologically-nontrivial band structure with ferromagnetism, which intrinsically breaks time-reversal symmetry (TRS), and therefore may produce unusual electromagnetic phenomena such as the topological magnetoelectric effect \cite{Qi2008,Essin2009}, quantized Kerr and Faraday rotation \cite{Tse2010,Wu2016}, and quantization of the anomalous Hall effect (QAHE) \cite{Yu2010, Chang2013}. However, while the quantum states in an ideal TI are protected by symmetry, most topologically-nontrivial materials have proven difficult to grow and engineer. In particular, residual bulk conductivity, and the tendency of TI materials to degrade during semiconductor processing, have impeded many experimental efforts \cite{Kong2011,Benia2011,Wang2012z}. Harnessing the unique physics of these materials will require better understanding and control of their magnetism and disorder, as well as methods to engineer TI devices without degrading the properties that make these materials unique. 

A major challenge in TI research is to identify systems in which the energies and symmetries of TI physics can be reliably controlled in the laboratory. For instance, a criterion for the QAH insulator state is for the chemical potential of electrons $\mu$ to lie within an energy gap $\Delta$ in the surface state dispersion, where $\Delta \propto |\vec{M}\cdot\hat{n}|$, $\vec{M}$ is the local magnetization, and $\hat{n}$ is the surface normal \cite{Hasan2010,Qi2011a}. At the edges of a uniformly magnetized film, the top and bottom surfaces meet and therefore $\vec{M}\cdot\hat{n}$ changes sign, causing $\Delta$ to vanish. In sufficiently thin films, the vanishing gap produces 1D conductive states that propagate along the edges of the film, and their quantized conductance is observed as the QAHE \cite{Yu2010,Chang2013}. Similar edge states are predicted to occur along magnetic domain walls, where the direction of magnetization inverts. These domain wall-bound states are of considerable present interest \cite{Wickles2012,Feng2015,Wang2015,Wang2015Retraction,Liu2016}, but they have not yet been observed directly. Methods to intentionally pattern the magnetization and chemical potential of TI materials \textit{in situ} during experiments may be particularly useful in efforts to detect, and eventually to harness, these unique phenomena.

Much research has focused on understanding and minimizing the sources of both magnetic \cite{Okada2011, Wang2015b, Lachman2015,Wang2016} and chemical potential \cite{Roushan2009, Beidenkopf2011, Kastl2015} disorder in TI materials. However, relatively few techniques have been explored to systematically control these parameters. Magnetic control of TI materials has focused largely on introducing TRS-breaking in different layers along the growth axis of thin films by modulation doping of magnetic ions \cite{Mogi2015}, or in magnetic-nonmagnetic bilayer devices \cite{Kou2013a}. Individual magnetic dopants have also been arranged on TI surfaces with STM techniques \cite{Schlenk2013}.  

Spatial control of chemical potential in TI materials is difficult due to material degradation during top gate fabrication and other cleanroom processes \cite{Kong2011,Benia2011,Wang2012z}. Previously, we reported an all-optical technique to tune the chemical potential of a variety of ultra-thin materials grown on \STO \cite{Yeats2015}. By optically manipulating space-charge in the underlying substrate, we created an effective local back-gate that was strong enough to create \textit{p}-\textit{n} junctions in non-magnetic thin films of \BST. These features were persistent and optically erasable, allowing arbitrary spatial control of the chemical potential without additional materials or fabrication. 

In this article, we demonstrate persistent, reconfigurable optical control over both the electron chemical potential and local magnetization of ferromagnetic thin films of Cr-\BST grown on \STO. First, we use simultaneous magneto-optical Kerr effect (MOKE) and anomalous Hall effect (AHE) measurements to establish a correspondence between optical and electrical measures of TI magnetism. We image the dynamics of magnetic coercion in the films using scanning Kerr microscopy, which suggest the presence of a nano-scale domain structure. We also employ both optical and electrostatic gating techniques to characterize both carrier-mediated and van Vleck-mediated ferromagnetism in the films. Finally, we demonstrate the ability to optically pattern and reconfigure the local chemical potential and local magnetization of the films using a two-color optical and magneto-optical recording protocol. We characterize these coexisting patterns with simultaneous scanning Kerr and photocurrent microscopies, showing that independent spatial control of both chemical potential and magnetization can be achieved at the same time in a ferromagnetic TI on a few-micron scale.

\section{Methods}
\label{sec:methods}

Samples consisted of 5-10 quintuple-layer (QL) films of Cr$_{x}$(Bi$_{y}$Sb$_{1-y}$)$_{2-x}$Te$_{3}$ grown by molecular beam epitaxy (MBE) on  (111)-oriented \STO.  Growth parameters and detailed material characterization data are presented in a separate manuscript \cite{Richardella2015}. The composition of the films is comparable to previous materials grown in the same MBE chamber that showed QAHE at very low temperatures ($T\ll1$ K) \cite{Liu2016,Lachman2015,Kandala2015}.  For consistency, we present data in the main text from one representative sample, a 9 QL film of Cr$_{0.15}$(Bi$_{0.5}$,Sb$_{0.5}$)$_{1.85}$Te$_3$ capped with a 4.5 nm layer of Al. Other films are discussed in the supplementary materials. Hall bars were patterned in all films by mechanically scratching away the growth layer with a needle, thereby avoiding sample contamination from lithographic processes.  The samples were mounted in an optical cryostat and an electromagnet was used to apply an out-of-plane magnetic field at the sample. 

The local magnetization of the film was imaged with scanning Kerr microscopy in the polar MOKE geometry. A beam of linearly polarized light was directed at normal incidence onto the sample and the rotation in polarization \TK of the reflected beam was measured using lock-in techniques as described in the supplementary materials. A reflective microscope objective was used to avoid unwanted Faraday rotation from glass optics \cite{Kondo2012}. Illumination was provided by a HeNe laser, with a photon energy ($\hbar \omega =$ 1.96 eV) in the vicinity of a known resonance in the MOKE spectrum of similar materials \cite{Patankar2015}. For simultaneous resistance, AHE, and MOKE hysteresis measurements, the laser spot was positioned on an unconnected leg of the Hall bar to avoid optical effects on the transport measurements. 

Resistance measurements were performed using standard AC lock-in techniques with an excitation current of 100 nA and a reference frequency of 17 Hz. Electrical contacts to all samples were made with indium and were ohmic. The chemical potential of the films was tuned by applying a voltage to the back-side of the substrate, or through a persistent, substrate-based optical gating effect \cite{Yeats2015}.  A fiber-coupled ultraviolet (UV, $\lambda =$ 370 nm) light emitting diode (LED)  was used to provide illumination for optical gating. For simultaneous Kerr and photocurrent imaging, a virtual null current preamplifier was used to measure the photocurrent induced by the MOKE laser spot as it was rastered across the sample.

\section{Results and Discussion}
\label{sec:results}

We used simultaneous magneto-optical and magneto-transport measurements to characterize the ferromagnetism of thin films of Cr-\BST at $T =$ 2.9 K. Fig. 1(a) shows the evolution of polar Kerr angle $\theta_{K}$ and anomalous Hall resistance \Rxy in a representative 9 QL film of Cr-\BST as a function of increasing and decreasing magnetic field. Clear, roughly square-shaped hysteresis is visible in both MOKE and AHE traces, indicative of ferromagnetic ordering with an easy axis out of the sample plane. We estimate the film has a Curie temperature $T_C \approx 18$ K (see supplementary materials). Fig. 1(b) shows the measurement geometry. Fig. 1(c) shows the longitudinal resistance \Rxx as a function of applied field at 2.9~K. The peaks at $\pm H_C$ have been observed in similar materials and are attributed to increased scattering from magnetic disorder during coercion \cite{Checkelsky2012}. Although similarly-composed materials grown in the same MBE chamber have shown quantized conductance at $T \ll 1 $ K \cite{Kandala2015,Lachman2015,Liu2016}, the absolute resistance values we observe indicate that conductivity is dominated by surface- and bulk-conductive channels (as opposed to quantized edge states) at the temperatures achievable in our optical cryostat ($T > 2.9$ K). 

Optical illumination during MOKE experiments could affect the magnetism of the films. Fig.~1(d) shows the coercive field $H_C$ extracted from MOKE hysteresis measurements as a function of laser power at $T =$ 3.2 K and with a spot size of $\approx$ 1 \um. The coercivity of the film decreases with illumination, fitting well to a power law relation (dashed line) from 20~\uW to 2~mW, where local heating is likely to dominate. The trend plateaus at low powers, showing good agreement with the coercivity extracted from AHE hysteresis in the absence of illumination (red arrow). This suggests that illumination from MOKE measurements conducted below $\approx$ 2 \uW are unlikely to significantly affect the magnetic properties of the film. 

\begin{figure}[] 
	\begin{center}
		\def\foo{Simultaneous MOKE  and AHE measurements}
		\includegraphics[width=8.6cm]{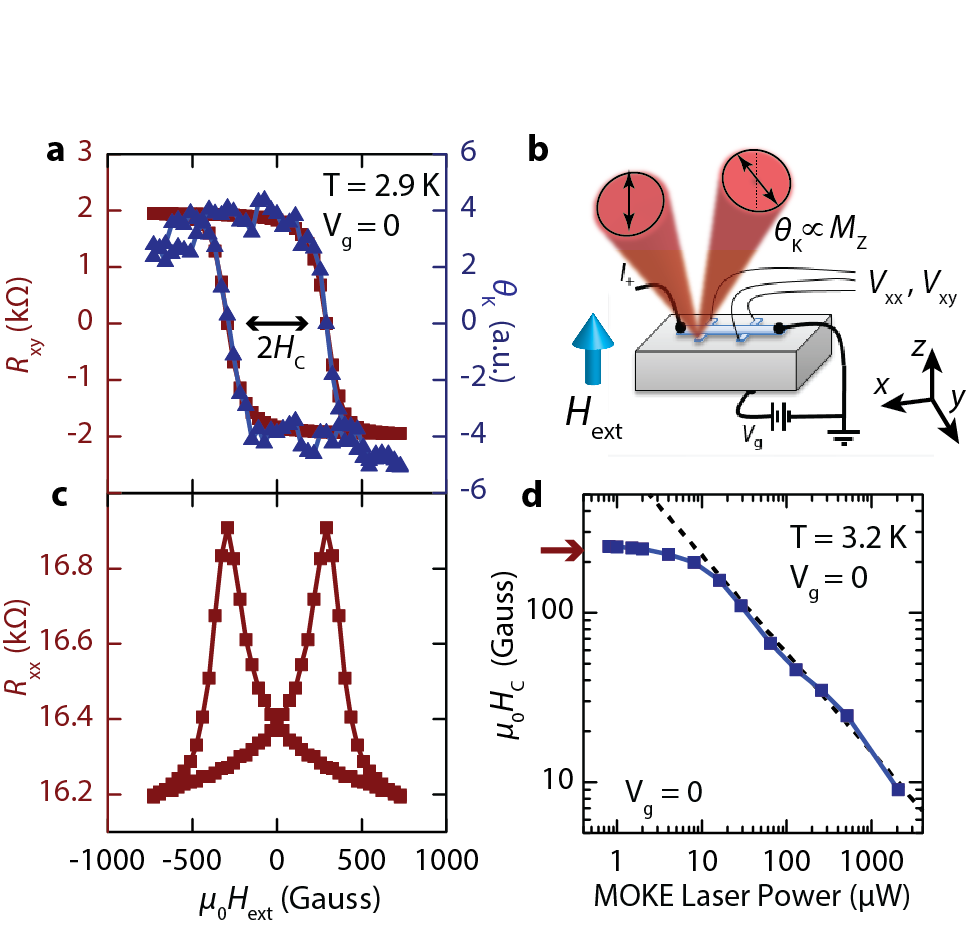}
		\caption[\foo]{\label{fig:Fig1} \textbf{\foo.}
			(a) Magnetic hysteresis loops showing anomalous Hall resistance \Rxy and Kerr angle $\theta_{K}$ as a function of magnetic field in a film of Cr-\BST at $T = 2.9$ K.  MOKE was conducted with a spot size of $\approx$ 1 \um and a laser power of  $\approx$ 1 \uW. 
			(b) Schematic of the measurement geometry. MOKE measurements were performed on the unused leg of the Hall bar to avoid optical effects on transport measurements. 
			(c) Magnetic hysteresis in longitudinal resistance \Rxx. 
			(d) Coercive field $H_C$ of the film from MOKE hysteresis measurements as a function of MOKE laser power. The red arrow indicates $H_C$ extracted from AHE hysteresis without illumination.
		}
	\end{center}
\end{figure}

By rastering the MOKE laser spot across the sample at low power, it is straightforward to produce spatial images of the magnetization of the films. Fig. 2 shows scanning Kerr micrographs of the reversal of remanent magnetization in a thin film of Cr-\BST after 5-s exposures to successively larger fields. The images show a continuous and spatially-uniform change in remanent magnetization. This is most likely an indication that magnetic inversion occurs on length scales smaller than the resolving power of our apparatus ($\approx$~1~\um). This is consistent with the sub-micron size domains observed by Lachman \textit{et al.} by scanning nano-SQUID microscopy in similar materials \cite{Lachman2015}. We see similar behavior in all of the films grown on \STO. By contrast, we observe inversion by domain-wall motion on longer length scales in one very heavily Cr-doped film grown on InP (see supplementary materials). However, this heavily-doped sample is unlikely to exhibit QAHE. 

\begin{figure}[] 
	\begin{center}
		\def\foo{Kerr microscopy of remanent magnetization}
		\includegraphics[width=8.6cm]{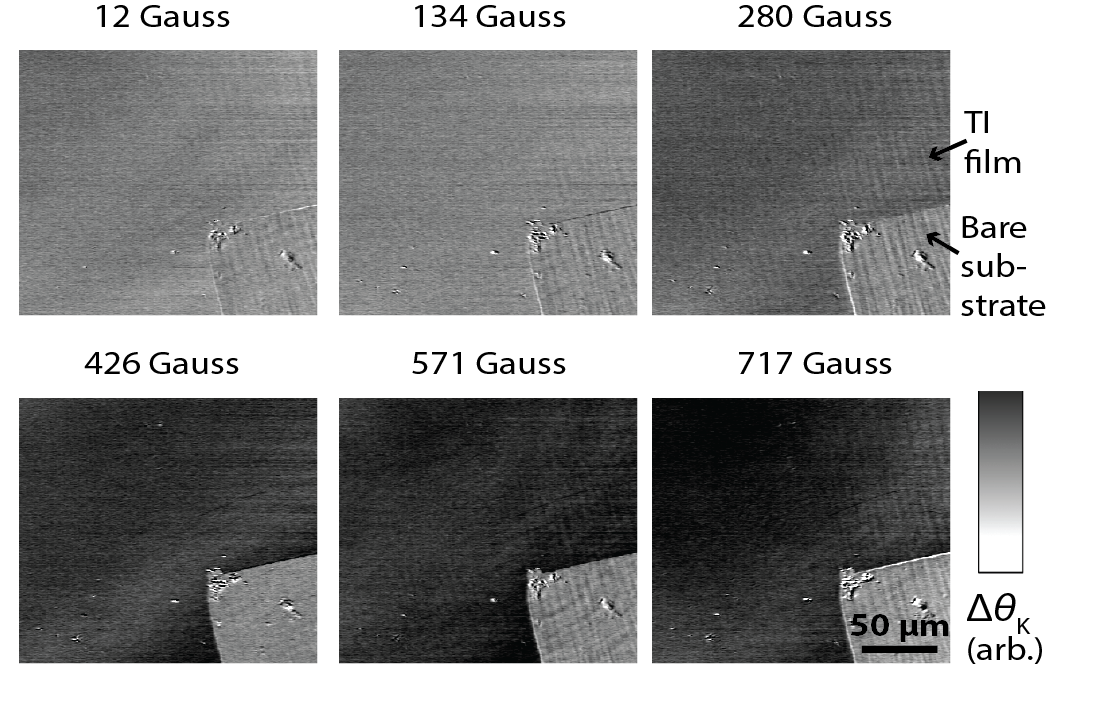}
		\caption[\foo]{\label{fig:Fig2} \textbf{\foo.}
			Zero-field Kerr microscopy showing the reversal of remanent magnetization  in a Cr-\BST thin film after exposure to successively larger opposing fields. The lower-right corner shows a region of bare substrate where the film was scratched away for reference. A zero-field baseline image (not shown) was subtracted from each image for clarity, such that the colorscale represents the change in remanent magnetization. The laser power was $\approx$ 2 \uW.
		}
	\end{center}
\end{figure}

\begin{figure}[] 
	\begin{center}
		\def\foo{Optical and electrostatic gate response}
		\includegraphics[width=8.6cm]{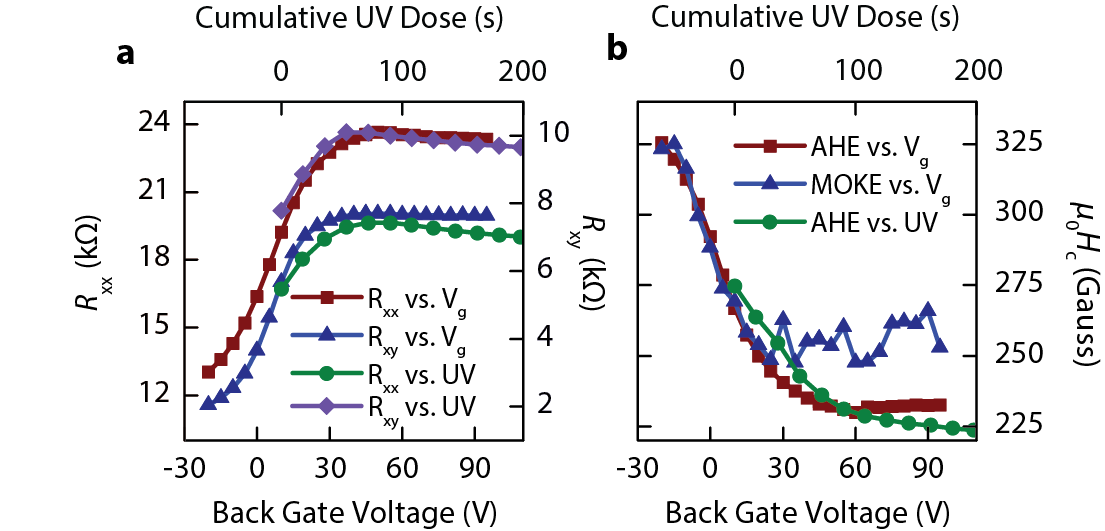}
		\caption[\foo]{\label{fig:Fig2} \textbf{\foo.}
			(a) Zero-field longitudinal resistance \Rxx and symmetrized Hall resistance \Rxy as a function of back gate voltage (red squares and blue triangles) or cumulative exposure to UV light (green circles and purple diamonds). Zeros are offset to account for hysteresis in the back gate response. The saturation fields for AHE measurements were $\mu_{0}H_{ext} = \pm 1000$ G.
			(b)  Coercive field $H_C$ extracted from AHE (red squares) and MOKE hysteresis (blue triangles)  measurements, plotted as a function of back-gate voltage (bottom axis). The green circles show $H_C$ extracted from AHE as a function of cumulative exposure to UV light (top axis). The UV intensity was $\approx$ 10 mW/m$^2$.
		}
	\end{center}
\end{figure}

Field-effect gating is generally needed to position the chemical potential within the surface state gap in QAHE studies \cite{Chang2013}. We used electrostatic back-gating, as well as a substrate-based persistent optical gating effect, to characterize the response of the film to the field effect. UV illumination produces a persistent field effect in thin films grown on \STO via electrical polarization of the underlying substrate, and this effect is functionally equivalent to a an electrostatic back-gate \cite{Yeats2015}. Fig. 3(a) shows the evolution of \Rxx and \Rxy as a function of either $V_g$ (bottom axis), or previous exposure to UV illumination (top axis).  The resistances evolve similarly with either type of gating. The response is consistent with expectations for an initially \textit{p}-type film that is gated just past its Dirac point to a weakly \textit{n}-type regime at $V_g \gtrsim 35$ V or $\gtrsim$~60~s of cumulative UV exposure. Because the optical gating effect is persistent, all measurements were performed in the dark $\approx$ 60 s after each exposure in order to avoid any transient effects from illumination of the sample.

Carrier-independent magnetism is important for magnetic TIs to achieve the QAHE because magnetization must remain after tuning the chemical potential to the surface-state energy gap \cite{Yu2010}. While carrier-mediated magnetism dominates in the more lightly Cr-doped films we have studied (see supplementary information), in the present film we observe magnetism that is largely independent of gate-tuning by either optical or electrostatic effects. Fig. 3(b) shows the evolution of $H_C$ (extracted from simultaneous AHE and MOKE measurements) as a function of electrostatic back-gating (bottom axis) or previous exposure to UV light (top axis). $H_C$ decreases by $\approx 30 \%$ with increasing $V_g$ (or previous exposure to UV light), consistent with only partially hole-mediated ferromagnetism. This is consistent with other reports of coexisting hole-mediated and van Vleck-mediated ferromagnetism in moderately Cr-doped \BST \cite{Kou2013,Richardella2015b}. Notably, the values of $H_C$ extracted from AHE and MOKE measurements correspond well in the absence of a gate effect, but at higher voltages (or UV exposures), MOKE hysteresis measurements indicate a larger $H_C$ than AHE hysteresis measurements. This may imply that optically-excited carriers from MOKE measurements can play a role in the carrier-mediated ferromagnetism of these materials, and this contribution is relatively more important when free holes are otherwise depleted by a back-gate.  

Many proposed applications of magnetic TI materials require spatial control of TRS-breaking. Fig. 4(a-c) shows a protocol for magneto-optical recording \cite{Chen1968} in Cr-\BST. After initialization with a field $H_{init} > H_C$, a high-intensity laser spot is used to locally reduce the coercivity of the film, such that a weak external field $-H_C < H_{record} < 0$ will invert the magnetization of the film only in the illuminated area. When the laser and field are shut off, the magnetic pattern remains, and can be imaged with low-power scanning Kerr microscopy. Figure 4(d) shows a Kerr micrograph of intentionally-patterned magnetization in Cr-\BST. The minimum feature size is $\approx 1$ \um. We find that magnetic patterns remain persistent $>$ 24 h unless they are erased with a coercive magnetic field or by raising the temperature above $T_C$. After erasure, no signs of the old pattern appear in subsequent Kerr images, and new patterns may be recorded. Fig. 4(e) shows magneto-optical recording as a function of temperature, indicating that the effect vanishes above $T_C$. Moreover, patterns were written across the edge of the film, and no patterning effect was observed on the bare substrate area. The lack of response in these areas rules out other explanations due to effects in the substrate, such as optical polarization of magnetic impurities \cite{Rice2014}.

\begin{figure}[] 
	\begin{center}
		\def\foo{Magneto-optical recording}
		\includegraphics[width=8.6cm]{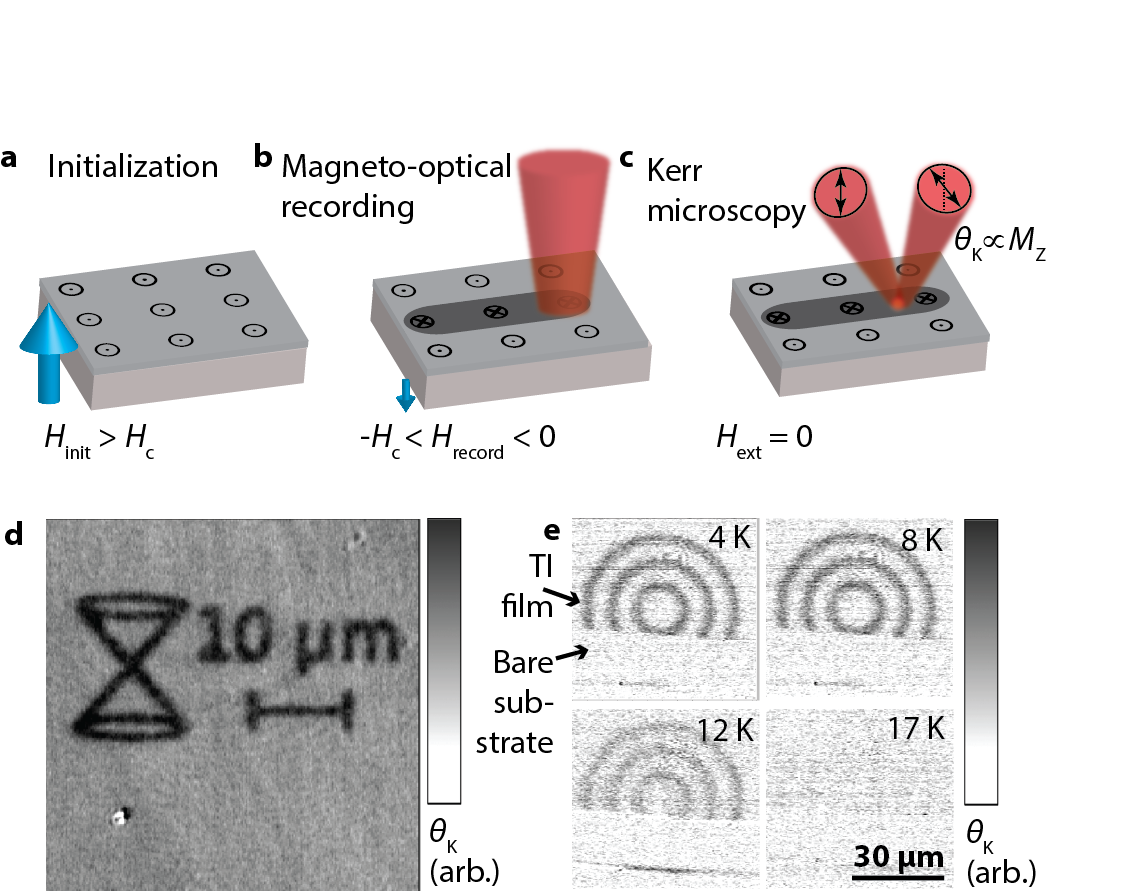}
		\caption[\foo]{\label{fig:Fig3} \textbf{\foo.}
			(a-c) Protocol for magneto-optical recording in a magnetic TI. (a) A strong magnetic field (1000 Gauss) is used to uniformly magnetize the film. (b) Laser illumination (32 \uW) locally reduces the coercivity of the film, allowing its magnetization to be inverted by a weak opposing field (-63 Gauss). (c) The pattern remains persistent at zero field and can be measured with scanning Kerr microscopy at low power (4 \uW). 
			(d) Scanning Kerr micrograph of magnetization pattern recorded in a thin film of Cr-\BST. 
			(e) Magneto-optical recording as a function of temperature. No effect is seen above $T_C$ or from the substrate in areas where the magnetic film has been scratched away.
		}
	\end{center}
\end{figure}

\begin{figure}[] 
	\begin{center}
		\def\foo{Simultaneous local control of magnetization and chemical potential}
		\includegraphics[width=8.6cm]{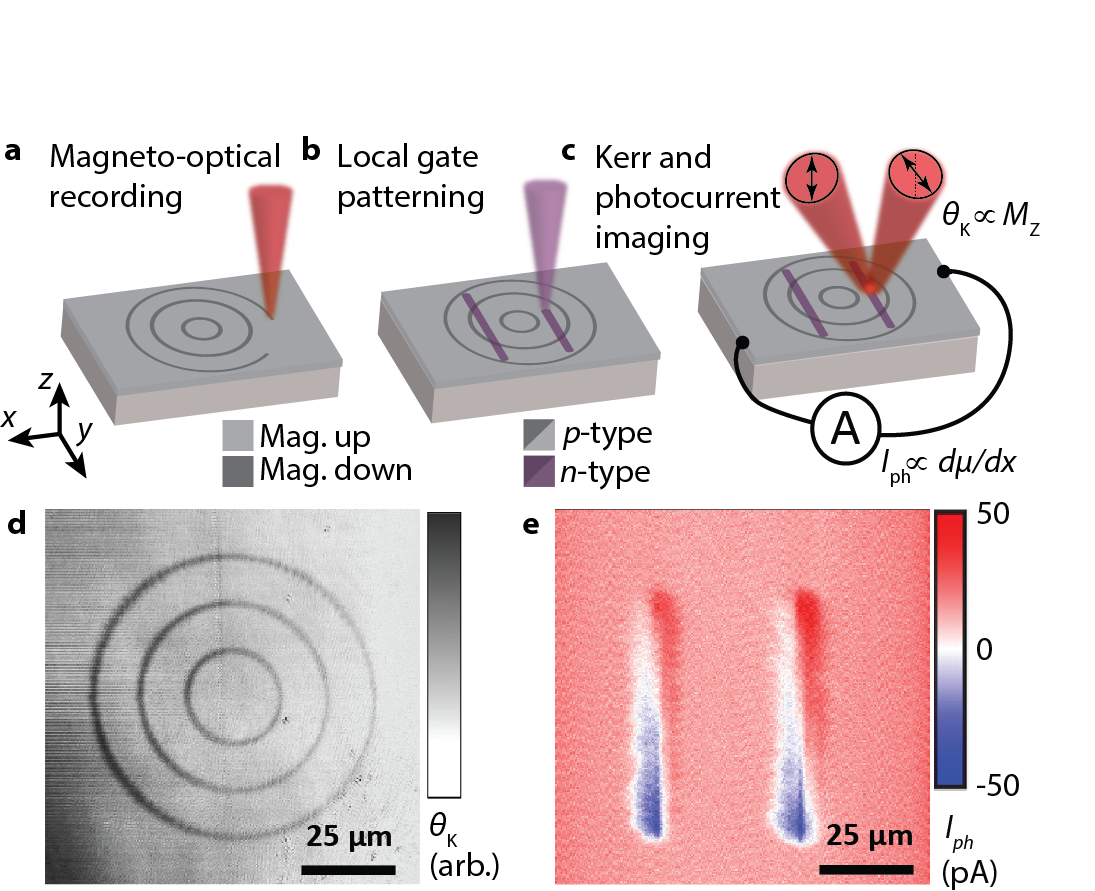}
		\caption[\foo]{\label{fig:Fig3} \textbf{\foo.}
			(a-c) Schematic of the patterning and imaging procedure.  (a) Three concentric circles are recorded in the magnetization of a Cr-\BST  film using the protocol shown in Fig. 4. (b) Focused UV illumination is used to locally increase the chemical potential of the film in the exposed areas, creating two \textit{n}-type regions in the otherwise \textit{p}-type film. (c) Scanning Kerr and photocurrent microscopy are performed simultaneously to measure both the local magnetization and local chemical potential gradient of the film. 
			(d) Scanning Kerr micrograph of three concentric circles recorded in the magnetization of a thin film of Cr-\BST.
			(e) Scanning photocurrent micrograph collected simultaneously with (d), showing the photoresponse of \textit{p}-\textit{n} junctions patterned in the same area. The image axes and pattern are tilted slightly with respect to the path between the two electrodes. The images were taken at zero field, $V_g = 0$,  $T = 3$ K. A relatively high laser power of 30 \uW was used in order to produce a clear photocurrent image.
		}
	\end{center}
\end{figure}

In addition to controlling the local magnetization through magneto-optical recording,  simultaneous patterning of the local chemical potential can be achieved by confining the substrate-based optical gating effect to a focused spot.  Fig. 5(a-c) shows a schematic of the patterning technique. First, magneto-optical recording is performed as described above to write three concentric circles in the local magnetization of the film. Then, two rectangular regions are exposed to UV light, which creates an effective local back-gate, raising the chemical potential of the film in those areas, and causing them to become \textit{n}-type. Scanning Kerr (Fig. 5(d)) and photocurrent (Fig. 5(e)) microscopies are then performed simultaneously using the same focused laser spot, as described in the methods. These images provide a map of the local magnetization and chemical potential gradient of the film, respectively. The photocurrent micrograph is consistent with the presence of local \textit{n}-type regions in the otherwise \textit{p}-type film. The image axes and pattern are tilted slightly with respect to the path between electrodes. The Kerr micrograph indicates that the magnetic pattern has survived the optical gating process, highlighting the presence of carrier-independent ferromagnetism in these films, and the ability of the films to support persistent, micron-scale magnetic features regardless of carrier density. 

Once written, both magnetic and chemical potential patterns persist indefinitely on accessible laboratory timescales. However, a relatively high laser power (30 \uW) is needed to produce clear photocurrent images, and so the imaging process relaxes the UV gating effect \cite{Yeats2015}. Repeated imaging can therefore be used to erase the field of view. The elevated laser power during magneto-optical recording may also relax the optical gating effect, and so magnetic recording must be performed as the first step in the process. Despite the elevated power during photocurrent imaging, the magnetic patterns do not appear to degrade so long as the external field is held at zero. Within these constraints, we find that magneto-optical recording and UV gate patterning combine well without noticeable cross-talk in the Kerr and photocurrent images. The combination of these two effects therefore allows independent, reconfigurable optical control of both the local magnetization (TRS-breaking) and chemical potential of a magnetic TI material on a few-micron scale.

\section{Summary and Outlook}
\label{sec:conclusion}

We have reported a set of optical measurement and control protocols, which we use to study and manipulate the local magnetization and chemical potential of thin films of Cr-\BST grown on \STO. In the low-power measurement regime, we show a close correspondence between MOKE and AHE measurements of magnetic hysteresis in the films, suggesting that polarization-resolved optical techniques complement electrical measurements in understanding the magnetic properties of TI films. We use scanning Kerr microscopy to image magnetic inversion in the films, suggesting the presence of a nano-scale domain structure. Furthermore, we use both MOKE and AHE measurements to investigate the partially carrier-mediated ferromagnetism of the films, which we then control either with an electrostatic back-gate or a persistent optical gating effect.  Finally by employing various scanning optical microscopy techniques, we demonstrate simultaneous local control of both the local chemical potential and local magnetization of the films. We show that both properties can be arbitrarily patterned, reconfigured, and imaged independently on a few-micron scale. 

Because the magnetization direction determines the sign of the surface state energy gap $\Delta$, the ability to control the magnetization of a Cr-\BST film at the micron scale may have immediate applications in efforts to detect the 1D edge states predicted to occur in TIs at magnetic domain wall boundaries \cite{Hasan2010,Qi2011a,Wickles2012}. Although the films we measure do not reach the QAH regime at 3 K, our approach is equally applicable in lower-temperature environments. Improvements in growth protocols, such as modulation doping of magnetic impurities \cite{Mogi2015}, may also bring QAH physics to the $T \approx 3$ K regime. Freedom to write and modify arbitrarily complicated magnetization patterns \textit{in situ} may be especially useful for prototyping devices based on TI physics, such as spin filters \cite{Hammer2013a} and spin-based transistors \cite{Hammer2013}. The magnetic patterning and microscopy techniques we present may also be useful in testing theoretical predictions about the mutual interactions between magnetic domain wall motion and spin-torque from TI surface states \cite{Tserkovnyak2012,Ferreiros2014}. Since both optical effects are persistent and bidirectional, a series of incremental recording steps may allow regions of a particular magnetization or chemical potential to be moved in two dimensions without changing topology. This might have relevance for studying solid state analogies to particle physics and related proposals for quantum information processing \cite{DasSarma2005,Nayak2008,Ferreira2013}.

\section{Acknowledgments}
\label{sec:acknowledgments}

We thank J. van Bree for useful discussions. We acknowledge use of the NSF National Nanofabrication Users Network Facility at Penn State. This work is supported by the Office of Naval Research (grants N00014-15-1-2369 and N00014-15-1-2370), the Air Force Office of Scientific Research Multidisciplinary University Research Initiative (grant FA9550-14-1-0231), the NSF Materials Research Science and Engineering Centers (grant NSF-DMR-1420709), and the National Science Foundation (grants NSF-DMR-1306300 and NSF-DMR-1306510). This study is based in part upon research conducted at The Pennsylvania State University Two-Dimensional Crystal Consortium – Materials Innovation Platform (2DCC-MIP) which is supported by NSF cooperative agreement DMR-1539916. This material is based upon work supported by Laboratory Directed Research and Development (LDRD) funding from Argonne National Laboratory, provided by the Director, Office of Science, of the U.S. Department of Energy under Contract No. DE-AC02-06CH11357.

\bibliographystyle{prxStyle}
\bibliography{library}

\end{document}